# Wintertime Cross-correlational Structures between Sea Surface Temperature Anomaly and Atmospheric-and-Oceanic Fields in the East/Japan Sea Under Arctic Oscillation

Gyuchang Lim[1], Jong-Jin Park[1,2]*

[1] Kyungpook Institute of Oceanography, Kyungpook National University, Daegu 41566, South Korea
[2] School of Earth System Sciences, Kyungpook National University, Daegu 41566, South Korea 2
* Correspondence: jjpark@knu.ac.kr

**Abstract:** The winter Arctic Oscillation (AO) modulates East Asian climate and the East/Japan Sea (EJS), yet local, scale-dependent air-sea couplings linking atmosphere, ocean and sea-surface temperature anomalies (SSTA) remain unclear. Using 30 years of daily fields (1993–2022), we compute detrended fluctuation/cross-correlation metrics over 5–50-day scales at every grid: the Hurst exponent ($H$), the cross-Hurst exponent ($h_{XY}$), and the DCCA coefficient ($\rho_{dcca}$). Significance is assessed with iterative-AAFT surrogates and Benjamini–Hochberg false-discovery-rate control. Three robust features emerge. (1) During AO+ winters, the EKB–SPF corridor exhibits high SSTA variance and near-ballistic persistence ($H \approx$ 1.4–1.5), indicating increased susceptibility to marine heatwaves. (2) SSTA co-fluctuates positively with near-surface air-temperature anomalies, whereas turbulent heat-flux anomalies are largely anti-phased and show negligible cross-persistence, consistent with fast damping. (3) Oceanic fields impart persistent coupling: sea-surface height anomalies display basin-wide positive links with SSTA; meridional geostrophic velocity imprints advective cross-coupling along EKWC/SPF pathways, while zonal flow and vorticity yield patchy signatures. Winter SST variability in the EJS thus reflects a two-tier process in which mesoscale structure and along-front advection organize persistence, while synoptic forcing and turbulent heat exchange supply strong but non-persistent tendencies. The FDR-controlled, grid-point DFA/DCCA framework is transferable to other marginal seas.

## 1. Introduction

The Arctic Oscillation (AO) is a leading mode of Northern Hemisphere variability that reorganizes the extratropical circulation and exerts a first-order control on East Asian winter climate. In its positive (negative) phase, a strengthened (weakened) polar vortex accompanies a pressure seesaw that generally weaken (strengthens) the East Asian winter monsoon, favoring warmer (colder) winters across Northeast Asia through adjustments in the Siberian High, Aleutian Low, and the East Asian jet stream [1–3]. These AO-forced circulation anomalies project onto near-surface winds and surface fluxes, providing a canonical pathway for AO influence on regional ocean–atmosphere exchanges over the marginal seas of East Asia [3].

The East/Japan Sea (EJS) is a semi–enclosed marginal sea bounded by Korea, Japan, and Russia. Its wintertime environment features a pronounced meridional sea surface temperature (SST) gradient with a climatological subpolar front (SPF) near ~40°N, persistent cold and dry north-westerlies under the monsoon flow, and vigorous turbulent heat exchange [4]. Observational syntheses over the East Asian marginal seas show large wintertime sensible and latent heat losses and strong event-scale variability, underscoring the need for high-frequency analysis of air-sea coupling [5]. In the EJS, SST variability is further organized by the Tsushima Warm Current (TWC)/East Korea Warm Current (EKWC) system and associated eddy activity, which modulate advective heat transport along the subpolar front and into the East Korea Bay (EKB) [4, 6–7].

Recent work has linked AO variability to EJS winter extremes, including marine heatwaves (MHWs). Using daily satellite SST and reanalyses, Song et al. [4] documented that during positive AO winters, anticyclonic eddy-like anomalies and Ekman downwelling develop around the EKB, coincident with anomalously warm SST and a marked increase in MHW days; notably, the anomalies could not be primarily explained by local net surface heat flux, pointing instead to oceanic dynamical adjustments as the proximal driver [4]. These results motivate a grid-point assessment of which specific atmospheric, oceanic, and coupled-flux fields co-vary with winter SSTA locally, and how such co-variability depends on AO phase.

From a process perspective, atmospheric fields can influence SST anomalies (SSTA) through both mechanical and thermodynamic pathways—momentum input (wind stress), wind-stress curl and Ekman pumping, and turbulent heat fluxes (sensible and latent) [4, 8]—while oceanic fields such as sea-surface height anomalies (SSHA) and geostrophic currents (geo-U and geo-V) directly impose advective and mesoscale controls on SST [8, 9]. Model-based studies over East Asia emphasize that realistic air-sea coupling at daily time scales requires representing wind-driven mixed-layer responses, surface roughness, and skin-temperature physics, especially when diagnosing feedbacks between SST and the lower troposphere [10]. These considerations suggest that different drivers may exhibit distinct scale-dependent couplings with SSTA—some resembling slow "integration" of synoptic atmospheric forcing by the mixed layer, others reflecting direct, advective ocean dynamics.

To quantify such scale-dependent persistence and cross-dependence between SSTA and atmospheric/oceanic/coupled forcing fields, we employ Detrended Fluctuation Analysis (DFA) [11] and Detrended Cross-Correlation Analysis (DCCA) [9,12,13], which characterize, respectively, the memory of a single timeseries (Hurst exponent, $H$) and the cross-persistence between two non-stationary timeseries (cross-Hurst exponent, $H_{XY}$) together with a scale-resolved DCCA coefficient, $\rho_{dcca}$

[14]. In the context of winter AO forcing, these methods are particularly attractive: (i) they accommodate non-stationarity and wideband variability inherent in daily anomaly fields; (ii) they can reveal integration-like responses expected for several atmospheric variables [15]; (iii) they also diagnose direct oceanic controls that may couple to SSTA without a slow atmospheric integration pathway [9]. Relative to our prior basin-scale study of wintertime EJS air-sea thermal interactions [16], which emphasized large-scale patterns and integration-time based air-sea coupling, we move here to grid-point diagnostics to resolve spatial heterogeneity and local couplings under opposite AO phases.

In this study, using 30 winters (1993–2022) of daily fields, we investigate, at each grid point of the EJS, the cross-persistence and scale-dependent cross-correlations between SSTA and (i) atmospheric anomalies (2m air temperature, sea-level pressure, wind-stress curl, and 10m zonal/meridional winds), (ii) coupled-flux anomalies (surface sensible and latent heat flux), and oceanic anomalies (sea-surface height and geostrophic currents/curl). Surrogate-based significance testing [17,18] and false discovery rate control [19] are applied to identify robust local couplings quantified by $H_{XY}$ and $\rho_{dcca}$. The remainder of this paper is organized as follows. Section 2 describes data and methods (DFA/DCCA workflow, surrogate generation, and multiple-testing controls). Section 3 presents the grid-point results for $H_{XY}$ and $\rho_{dcca}$ across atmospheric, oceanic, and coupled fields under positive vs. negative AO. Section 4 discusses the physical interpretations—persistence vs. direct dynamical control—and consequences for winter MHW risk and prediction over the EJS.

## 2. Materials and Methods

### 2.1 Data

#### 2.1.1 Sea Surface Temperature (SST)

Daily Optimum Interpolation SST, version2.1 (OISST v2.1), from the National Oceanic and Atmospheric Administration (NOAA) was used on a 0.25° × 0.25° grid [20,21]. OISST blends AVHRR satellite retrievals with in-situ ship, buoy and Argo observations, applies bias correction to drifting/buoy records, and includes sea-ice adjustments in high-latitude grid cells [20]. We analyzed the EJS (34°–45° N, 127°–144° E) for 1 January 1993–31 December 2022. SSTA were formed by removing a day-of-year climatology computed over 1993–2022 from the daily fields at each grid point.

#### 2.1.2 Atmospheric and air-sea-coupled variables (ERA5)

Atmospheric and surface-flux fields were taken from the ERA5 reanalysis produced by ECMWF [22]. We used single-level variable on the native 0.25° grid, retrieved 6-hourly (00, 06, 12, and 18 UTC), then averaged to daily means and bilinearly mapped onto the OISST grid. The atmospheric set comprises: 2m air temperature (ATMP), sea-level pressure (SLP), 10m zonal and meridional winds (U10, V10), and turbulent surface stresses ($\tau_x$, $\tau_y$). Wind-stress curl (CurlTau) was computed from the ERA5 stress components, not from U10/V10 directly, using second-order centered differences:

$$\text{Curl } \tau = \frac{\partial \tau_y}{\partial x} - \frac{\partial \tau_x}{\partial y}. \tag{1}$$

Daily anomalies (ATMPA, SLPA, UA10, VA10, CurlTauA) were obtained by subtracting day-of-year climatology (1993–2022), respectively.

For the air-sea coupled variables, we used ERA5 surface sensible heat flux (SSHF) and surface latent heat flux (SLHF), which are archived as time-integrated accumulation ($Jm^{-2}$). Following ECMWF guidance, we de-accumulated each 6-hourly step to flux units ($Wm^{-2}$) and then formed daily means. Let $A(t_i)$ be the accumulated value at valid times , $t_i \in \{06,12,18,24 \text{ UTC}\}$. The 6-hourly mean flux is

$$F_{6h}(t_i) = \frac{A(t_i) - A(t_{i-1})}{\Delta t}, \quad \Delta t = 6 \times 3600 s, \tag{2}$$

and the daily mean is

$$\bar{F}_{day} = \frac{1}{4} \sum_{k=1}^{4} F_{6h,k}. \tag{3}$$

We retain the ECMWF sign convention (positive downward into the surface). Thus, over the ocean, upward turbulent heat loss from the sea to the atmosphere appears as negative flux; anomalies in our maps preserve this sign.

### 2.1.3 Oceanic Variables

Sea-surface height anomaly (SSHA) and the associated geostrophic currents were taken from the DUACS (Data Unification and Altimeter Combination System) delayed-time, multi-mission gridded products distributed by the Copernicus Marine Environment Monitoring Service (CMEMS) at 0.25° resolution, daily, from 1 January 1993 onward [23,24]. Horizontal geostrophic velocities were provided by CMEMS and are derived from the mapped sea-level fields via the standard geostrophic relation; we denote them as geo-U and geo-V. The geostrophic-current curl (geo-Curl) was computed with second-order centered differences,

$$\text{geo-Curl} = \frac{\partial \text{geo-}V}{\partial x} - \frac{\partial \text{geo-}U}{\partial y}, \tag{4}$$

on the CMEMS grid; to avoid numerical edge effects, we masked one grid ring adjacent to land before differencing. Daily anomalies (SSHA, geo-UA, geo-VA, geo-CurlA) were formed by removing a 1993–2022 day-of-year climatology at each grid point.

### 2.1.4 Arctic Oscillation (AO) & phase selection

The AO index was obtained from the NOAA Climate Prediction Center (CPC) [25], which projects 1000 hPa height anomalies onto the leading empirical mode of Northern Hemisphere variability (cf. the AO framework of Thompson & Wallace [26]). For each year, we computed the JMF (January–February–March)-mean AO. Winters with JMF AO $> +0.8\sigma$ ($\sigma$ computed from the 1993–2022 JFM series) were tagged AO+, and those with AO $> -0.8\sigma$ were tagged AO−; remaining winters were not used for the phase-contrasted compositing. For all fields, we first extracted JMF segments, concatenated them ("cut-and-stitch"), and then applied the anomaly definitions above before the DFA/DCCA analyses.

## 2.2 Methodology

We diagnose scale-dependent persistence and cross-dependence between SSTA and candidate drivers using DFA and DCCA. DFA estimates a (univariate) Hurst exponent $H$ that summarizes long-range memory in non-stationary series [11,27]. DCCA extends DFA to pairs of series, yielding a cross-Hurst exponent $H_{XY}$ that characterizes the persistence of their detrended covariance, and a normalized, scale-resolved cross-correlation coefficient $\rho_{dcca}(s) \in [-1, 1]$ [12,14,18]. In the context of winter air-sea interactions, these tools are appropriate because (i) daily anomalies are broadband and weakly nonstationary, (ii) several atmospheric influences on SST (e.g., pressure and winds acting through mixed-layer integration) can yield persistent responses even when instantaneous Pearson correlations are small [15,28], and (iii) oceanic fields (SSH, currents, vorticity) can imprint direct, advective controls that act at synoptic and mesoscale time scales. All analyses are performed grid-point-wise and separately for the AO+ and AO− winter subsets.

### 2.2.1 Detrended Cross-Correlation Analysis (DCCA)

Let $\{x_i, y_i\}_{i=1}^N$ be a pair of daily anomaly series at a grid cell. Define the cumulative profiles

$$X(k) = \sum_{i=1}^{k} [x_i - \langle x \rangle], \qquad Y(k) = \sum_{i=1}^{k} [y_i - \langle y \rangle] \qquad (5)$$

where $\langle \cdot \rangle$ denotes the temporal mean. For a given scale $s$ (here, $5 \leq s \leq 50$ days), partition each profile into $N_s = \text{int}(N/s)$ non-overlapping segments from the beginning and $N_s$ from the end (total $2N_s$ segments). In each segment $v$, remove a local polynomial trend of order $m$ (we use $m = 1$) from both profiles, producing $\tilde{X}_v^{(m)}(k)$ and $\tilde{Y}_v^{(m)}(k)$. The detrended covariance in segment $v$ is

$$F_{XY}^2(v, s) = \frac{1}{s} \sum_{k=1}^{s} \left\{ \left( X[(v-1)s + k] - \tilde{X}_v^{(m)}(k) \right) \times \left( Y[(v-1)s + k] - \tilde{Y}_v^{(m)}(k) \right) \right\} \qquad (6)$$

Average across segments to obtain the fluctuation function

$$F_{XY}(s) = \left[ \frac{1}{2N_s} \sum_{v=1}^{2N_s} F_{XY}^2(v, s) \right]^{1/2}. \qquad (7)$$

If a power-law holds over a pre-specified scale band $S$ (here, 5–50 days), then

$$F_{XY}(s) \sim s^{H_{XY}}, \qquad s \in S, \qquad (8)$$

and the slope $H_{XY}$ (ordinary-least-squares regression of $\log F_{XY}$ on $\log s$) quantifies cross-persistence in the detrended covariance [12,14]. For identical series, DCCA reduces to DFA and $H_{XY} = H$ [11]. For heterogeneous series (persistent vs. anti-persistent) driven by a common white-noise forcing, $H_{XY}$ tends toward 0.5 [29,30].

To resolve the sign and magnitude of cross-dependence at each scale, we also compute Zebende's DCCA coefficient

$$\rho_{dcca}(s) = \frac{F_{XY}^2(s)}{F_X(s)F_Y(s)} \in [-1, 1] \quad (9)$$

where $F_X(s)$ and $F_Y(s)$ are DFA fluctuation function of the two series at the same detrending order [14]. We evaluate $\rho_{dcca}(s)$ for all $s \in S$ and, for display, also form a simple average across $S$ to obtain a scale-mean $\bar{\rho}_{dcca}$ map (significance is assessed per-scale; see Section 2.2.3).

### 2.2.2 Monte-Carlo significance and quality control for $H_{XY}$ and $\rho_{dcca}$

There are no closed-form null distributions for $H_{XY}$ or $\rho_{dcca}(s)$. We therefore adopt a surrogate-based Monte-Carlo framework following Podobnik et al [18]. For each cell $i$ and variable pair $(x, y)$, we generate $N_{sur} = 1000$ iterative amplitude-adjusted Fourier transform (iAAFT) surrogate pairs $\{x_{i,sur}^{(k)}, y_{i,sur}^{(k)}\}_{k=1}^{N_{sur}}$ that preserve the empirical amplitude distributions and approximately match the power spectra (hence the linear autocorrelations) of the originals [17]. Two-sided Monte-Carlo p-values are

$$p_\rho(i, s) = \frac{\#\{k: |\rho_{sur}^{(k)}(i, s)| \geq |\rho_{obs}(i, s)|\}}{N_{sur}}$$

$$p_H(i) = \frac{\#\{k: |H_{sur}^{(k)}(i) - 1| \geq |H_{obs}(i) - 1|\}}{N_{sur}} \quad (10)$$

For $H_{XY}$, the log–log linearity implicit in Eq. (8) is verified by the coefficient of determination $R^2$ of the fit over $S$. We pre-declare a quality threshold $R^2 \geq R_0$ (here, $R_0 = 0.90$). Cells not meeting this criterion are labeled "no reliable cross-persistence" and masked in $H_{XY}$ figures rather than being assigned $H_{XY} = 1$. The same check is applied to each surrogate. Let $K_i = \{k: R_{sur}^{2,(k)}(i) \geq R_0\}$ and $N_{\text{eff}}(i) = \#K_i$. We then compute

$$p_H(i) = \frac{\#\{k \in K_i: |H_{sur}^{(k)}(i) - 1| \geq |H_{obs}(i) - 1|\}}{N_{\text{eff}}(i)}. \quad (11)$$

If $N_{\text{eff}}(i) < 200$ (20% of $N_{sur}$), the cell is deemed not testable for $H_{XY}$ and is masked. This treatment avoids anticonservative p-values that would arise from coercing poorly fit surrogates to $H = 1$.

### 2.2.3 False-discovery-rate (FDR) control across space and scale

Because we test thousands of hypotheses in parallel (across grid cells, and for $\rho_{dcca}$ also across scales), we control multiplicity using the Benjamini–Hochberg (BH) procedure at target FDR $q = 0.05$ [19]. Families are defined as follows:

(1) For $\rho_{dcca}(s)$, for each fixed scale $s$, the family comprises all cells' p-values $\{p_\rho(i, s)\}$.

(2) For $H_{XY}$, there is one statistic per cell, so the family is $\{p_H(i)\}$ over all cells.

Within each family (and for each predictor–SSTA pair), sort p-values $p_{(1)} \leq \cdots \leq p_{(m)}$ and find $k^* = \max\{k: p_{(k)} \leq (k/m)q\}$. Declare the first $k^*$ tests significant at FDR $q$; if no such $k$ exists, no cell is significant.

For visualization, we optionally convert p-values to BH $q$-values (the minimal FDR level at which each test would be called significant) and mask cells that are not significant. When scale-mean $\bar{\rho}_{dcca}$ maps are shown, blank cells indicate locations where $\rho_{dcca}(s)$ failed BH at all scales in $S$ (or where fewer than a minimal fraction of scales passed, as noted in captions).

### 2.2.4 Implementation details and robustness

All variables are bilinearly regridded to the OISST 0.25° grid before analysis. We use detrending ($m = 1$) for DFA/DCCA, integer scales $s = 5, \cdots, 50$ days, and ordinary-least-squares regression on $\log s$ to estimate slopes. Results are qualitatively insensitive to using a slightly narrow scale band (e.g., 7–45 days) or to exclude the smallest two scales; these checks alter magnitude modestly but not the spatial patterns emphasized here. For $\rho_{dcca}$, we report both per-scale maps and (for compactness) scale-means $\bar{\rho}_{dcca}$ over $S$; significance is always assessed at the per-scale level as described above.

## 3. Analysis Results

This section examines how the wintertime AO phase modulates local air-sea coupling in the EJS by mapping (i) DCCA-based cross-persistence ($H_{XY}$) and (ii) scale-dependent detrended cross-correlations $\rho_{dcca}(s)$ between SSTA and three classes of drivers: atmospheric anomalies (ATMPA, SLPA, UA10, VA10, CurlTauA), coupled turbulent heat-flux anomalies (SSHFA, SLHFA), and oceanic anomalies (SSHA, geo-CurlA, geo-UA, geo-VA). Synoptic atmospheric "weather" typically forces SST via mixed-layer integration, so air-to-sea impacts build over multi-day to multi-week windows; by contrast, turbulent heat fluxes act as fast, largely damping feedbacks, while ocean currents imprint advective tendencies on intermediate time scales. The DCCA framework separates persistence (via $H_{XY}$) from phase-coherent co-variability across scales (via $\rho_{dcca}(s)$).

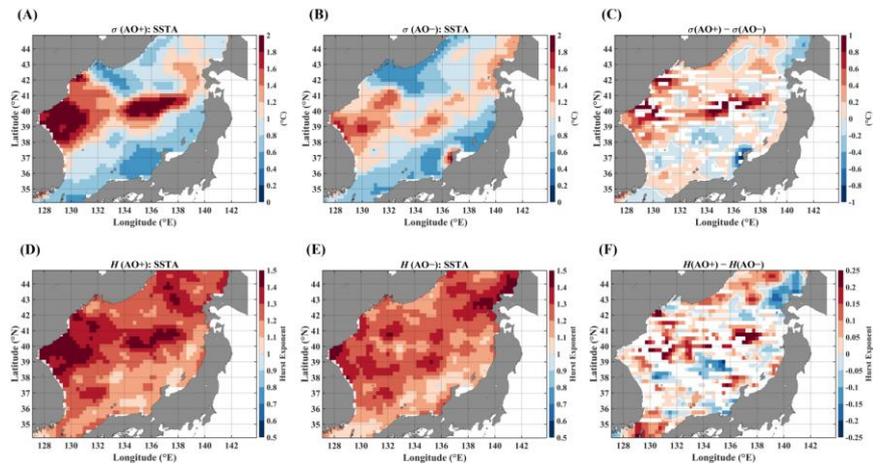

**Figure 1**. (A–C) Wintertime SSTA standard deviation (STD) for positive AO, negative AO, and "AO+ minus AO−" difference (1993–2022). (D–F) DFA-based Hurst exponent $H$ of SSTA for positive AO, negative AO, and their difference. Blanks in (C) and (F) mark insignificance at the 95% level.

### 3.1. SSTA variability and persistence: AO–phase contrasts

Wintertime SSTA variance (Figure 1A–C) is maximized in the East Korean Bay (EKB) and along the Subpolar Front (SPF). The AO-positive composite shows markedly higher standard deviation along the SPF and in the western basin relative to AO−, yielding a positive "AO+ minus AO−" difference in those corridors (Figure 1C). DFA-based persistence (Figure 1D–F) co-locates with these variance hot spots: during AO+, Hurst exponents are broadly elevated ($H \approx 1.2$–$1.5$), indicating strong long-range memory in daily SSTA. This co-occurrence of large variance and high $H$ under AO+ is consistent with a heightened propensity for winter marine heatwaves (MHWs) in the EKB/SPF sector, even though the full causal chain is beyond the present scope. These baseline contrasts identify the regions where subsequent DCCA diagnostics merit closest attention.

### 3.2. Local coupling with atmospheric fields

In all atmospheric-coupling figures, panels (A, B) show the cross-persistence $H_{XY}$ for AO+ and AO−, respectively; panels (C, D) show the scale-averaged $\bar{\rho}_{dcca}$ over $s \in [5, 50]$ days for AO+ and AO−. Cells are blank where $H_{XY}$ is not reliably defined (log–log fit $R^2 < R_0$) or fails BH-FDR at $q = 0.05$ (Section 2.2.3). For $\rho_{dcca}$, a cell is blank if no scale within $s \in [5, 50]$ survives BH-FDR.

**SSTA vs. ATMPA (Figure 2)**: Basin-wide positive $\bar{\rho}_{dcca}$ dominates in both AO phases, with greater spatial extent and amplitude during AO−. Thus, over 5–50-day windows, warmer air tends to co-vary locally with warmer SST (and vice versa), consistent with mixed-layer integration [15,16,28]. Significant $H_{XY}$ appears in elongated patches along the SPF/EKB and coastal bands—more continuous in AO−—indicating that persistent, scale-invariant co-evolution emerges where mixed-layer memory is strong and air-sea coupling is sustained.

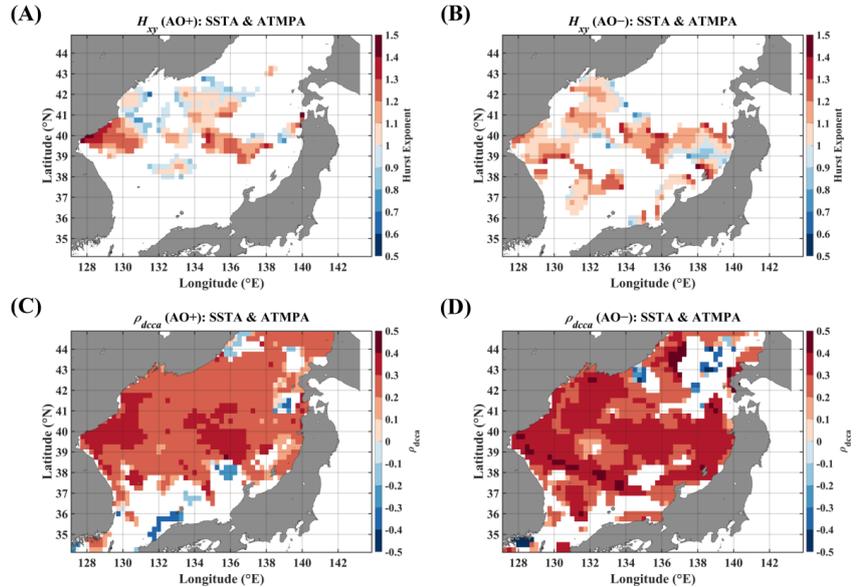

**Figure 2.** SSTA↔ATMPA. (A, B) DCCA cross-persistence $H_{XY}$ during AO+ and AO−. Cells are shown only where the log–log fit of $F_{XY}(s)$ vs. $s$ over 5–50 days meets the pre-declared quality criterion (e.g., $R^2 \geq R_0$) and the iAAFT Monte-Carlo p-value is BH-significant at FDR $q = 0.05$ (family: all grid cells). (C, D) Scale-averaged $\bar{\rho}_{dcca}$ (mean over 5–50 days). A cell is colored only if, for at least

one scale in the band, the two-sided surrogate $p_\rho(s)$ is BH-significant at $q = 0.05$ (family: all cells at fixed $s$); otherwise, it is left blank.

**SSTA vs. CurlTAuA (Figure 3)**: $\bar\rho_{dcca}$ forms a red/blue mosaic that reorganizes between AO phases (AO− favors broader negative belts in the central basin). Significant $H_{XY}$ pixels are sparse. This is consistent with wind-stress curl driving intermittent vertical motions and mixing whose sign and timing depend on storm-track geometry; such effects project onto SSTA within sub-windows of 5–50 days but rarely sustain cross-persistence.

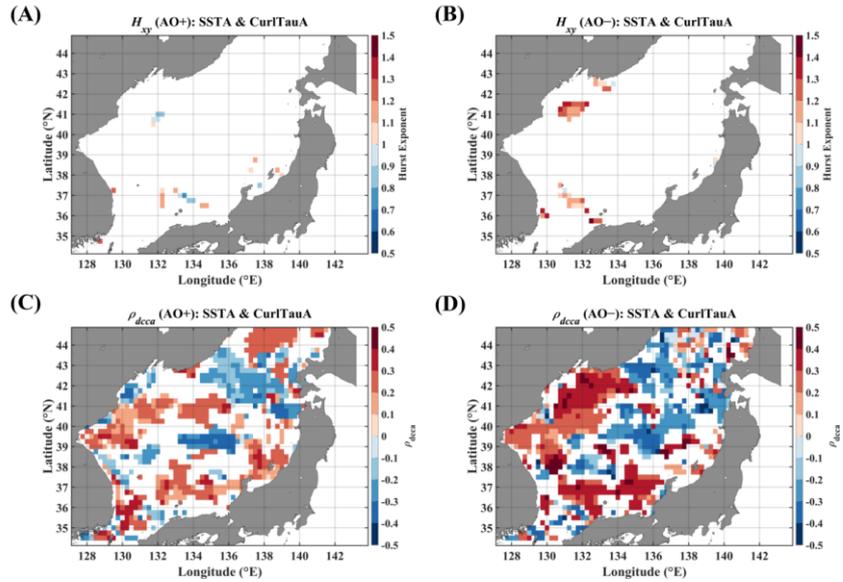

Figure 3. SSTA↔CurlTauA. Same layout and significance rules as Figure 2.

**SSTA vs. SLPA (Figure 4)**: Cross-persistence is largely absent, and $\bar\rho_{dcca}$ is regionally selective (e.g., positive belts near the northern margin in AO+ with mixed patterns in AO−). As SLP organizes synoptic systems rather than acting as a local flux, it modulates the placement and sign of weather forcing without yielding a persistent, scale-invariant local coupling with SSTA.

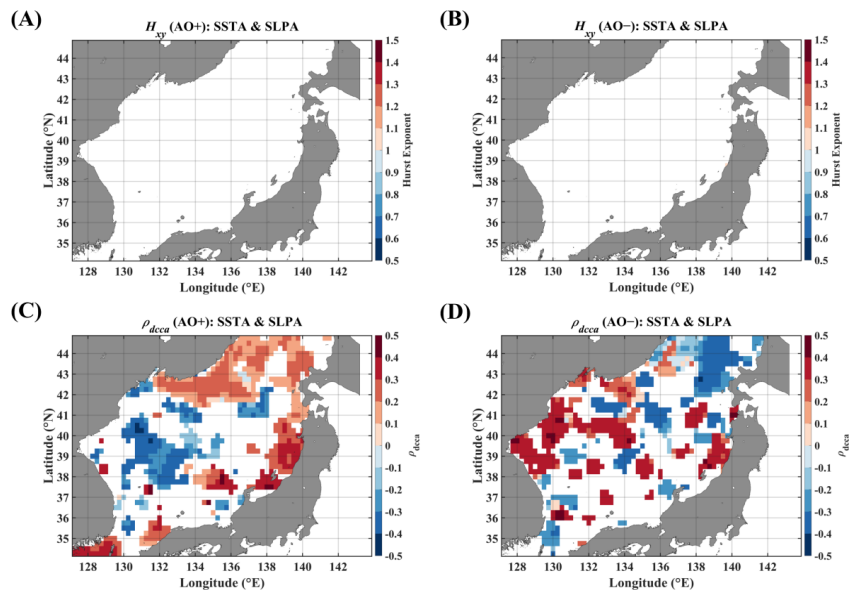

**SSTA vs. UA10 (Figure 5)**: $\bar{\rho}_{dcca}$ is predominantly negative in both AO phases, stronger and more extensive under AO−. Westerly anomalies enhance surface cooling and offshore Ekman transport in winter, favoring colder SST; easterlies produce the opposite tendency. Despite this robust sign, $H_{XY}$ is largely undefined, consistent with rapidly varying synoptic winds that impose tendencies without establishing cross-memory.

**SSTA vs. VA10 (Figure 6)**: $\bar{\rho}_{dcca}$ is mostly positive, with particularly strong AO− signals across the southern–central basin. Significant $H_{XY}$ appears more frequently under AO−, especially along the EKB/SPF corridor, suggesting that southerly anomalies (warm-advection-like patterns) are more phase-coherent with SSTA during cold AO winters when meridional pressure gradients are stronger. Under AO+, positive $\bar{\rho}_{dcca}$ persist but cross-persistence weakens, implying shorter, event-like coupling.

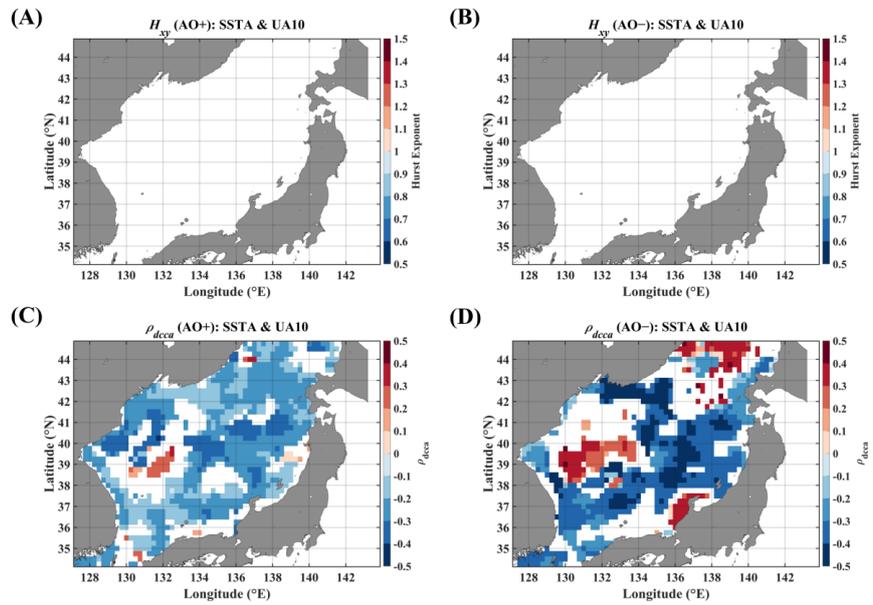

**Figure 5.** SSTA↔UA10. Same layout and significance rules as Figure 2.

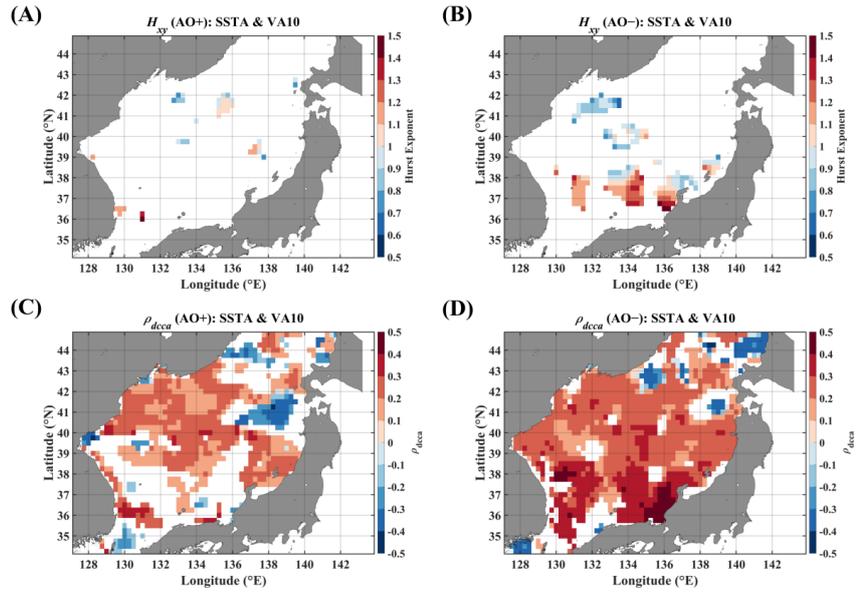

**Figure 6.** SSTA↔VA10. Same layout and significance rules as Figure 2.

Summarizing, across 5–50 days, the sign of local coupling is robust (positive for ATMPA and VA10, negative for UA10), whereas cross-persistence is selective—emerging mainly where mixed-layer memory is large (EKB/SPF) and under AO− configurations that reinforce advection/integration. Wind-stress curl and SLP imprint patchy $\rho_{dcca}$ with little persistence, consistent with their intermittent, geometry-dependent impacts.

### 3.3. Local coupling with coupled heat-flux anomalies

For both SSHFA (Figure 7) and SLHFA (Figure 8), $\bar{\rho}_{dcca}$ is predominantly negative across the basin in both AO phases—the canonical negative feedback: positive SSTA drives upward turbulent heat loss (cooling), whereas negative SSTA reduces the loss (warming). Crucially, $H_{XY}$ is virtually absent, indicating that—even when strong in sign—the flux–SSTA coupling is fast and dissipative, leaving no scale-invariant cross-memory over 5–50 days. AO primarily modulates the areal extent of significant negative $\bar{\rho}_{dcca}$, with AO− showing broader coverage in the interior basin, consistent with frequent cold-air outbreaks.

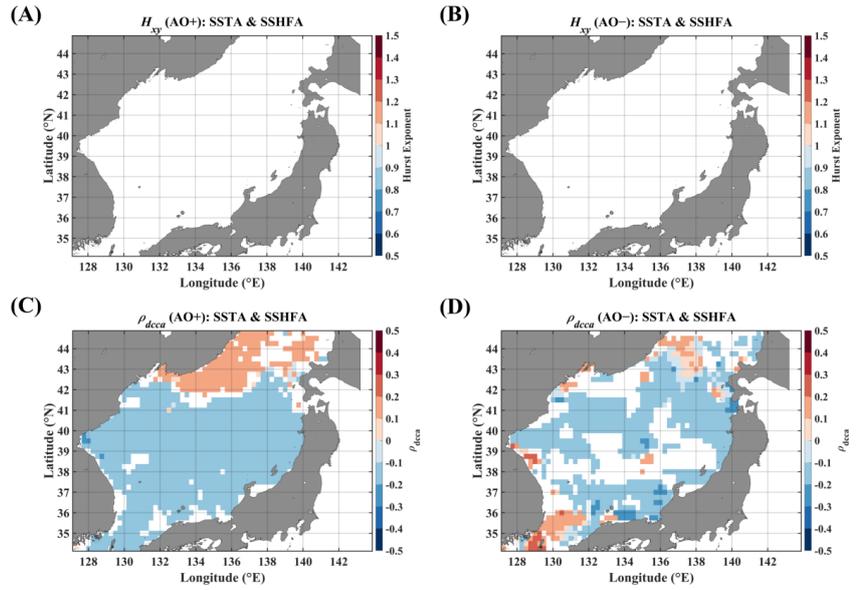

**Figure 7. SSTA↔SSHFA**. Same layout and significance rules as Figure 2. Negative $\bar{\rho}_{dcca}$ indicates the expected damping response of the ocean to positive SSTA.

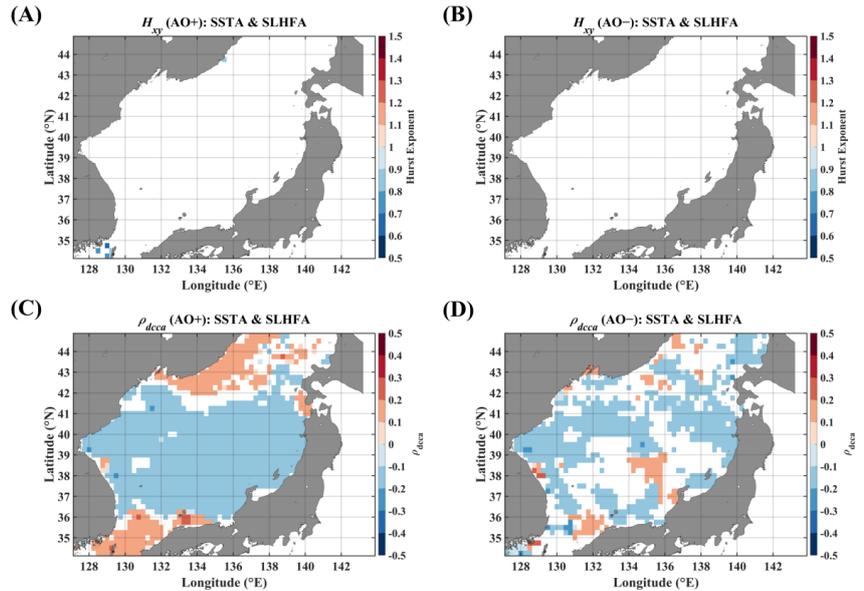

**Figure 8. SSTA↔SLHFA**. Same layout and significance rules as Figure 2. The basin-wide negative $\bar{\rho}_{dcca}$ reflects the canonical turbulent-flux feedback in winter.

### 3.4. Local coupling with oceanic fields

Mesoscale sea-level and geostrophic-current anomalies provide the oceanic pathways through which temperature is advected and retained in winter. Because SSHA integrates steric and dynamical signals, a positive association with SSTA is expected (warm/thick anomalies raise sea level), whereas velocity anomalies influence SSTA via along-front advection and eddy stirring. The DCCA maps confirm these expectations but also reveal clear AO modulations.

**SSTA vs. SSHA (Figure 9)**: $\bar{\rho}_{dcca}$ is basin-wide positive in both AO phases, with especially coherent coverage in AO− across the western and central basin. Patches of significant $H_{XY}$ occur along/near the SPF/EKB corridor and in the northern interior, indicating shared, scale-invariant co-memory where mesoscale features repeatedly affect the same locations. Among all drivers, SSHA exhibits the most extensive and AO-robust positive coupling, underscoring the central role of mesoscale structure in organizing SSTA variability.

**SSTA vs. geo-CurlA (Figure 10)**: A fine-scale red/blue mosaic appears in both phases with little $H_{XY}$. Horizontal vorticity primarily affects SSTA via stirring and deformation, producing intermittent, sign-changing local tendencies that are highly sensitive to frontal geometry; small spatial shifts of eddies/fronts can flip the local sign.

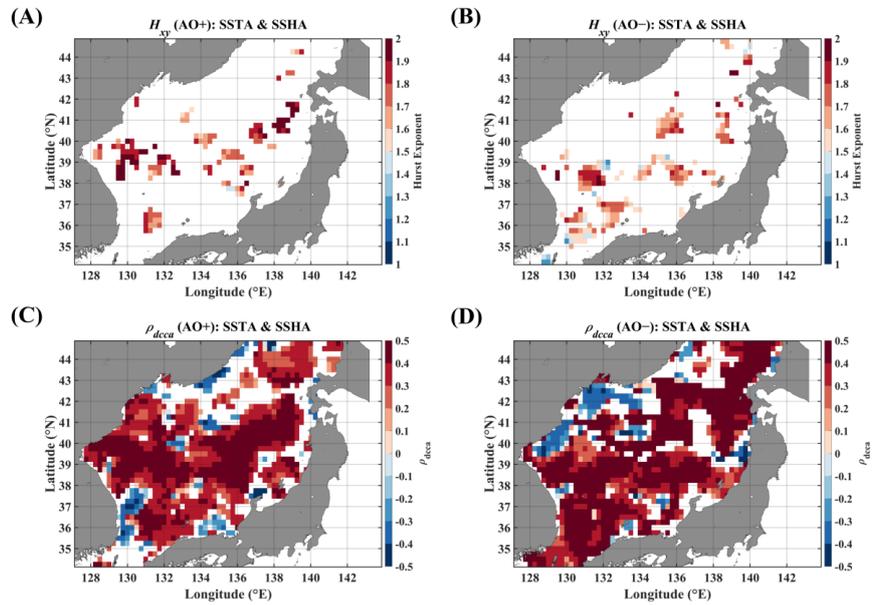

**Figure 9. SSTA↔SSHA.** Same layout and significance rules as Figure 2. White indicates non-significant or undefined cells. DUACS SSHA fields are used to characterize mesoscale structure.

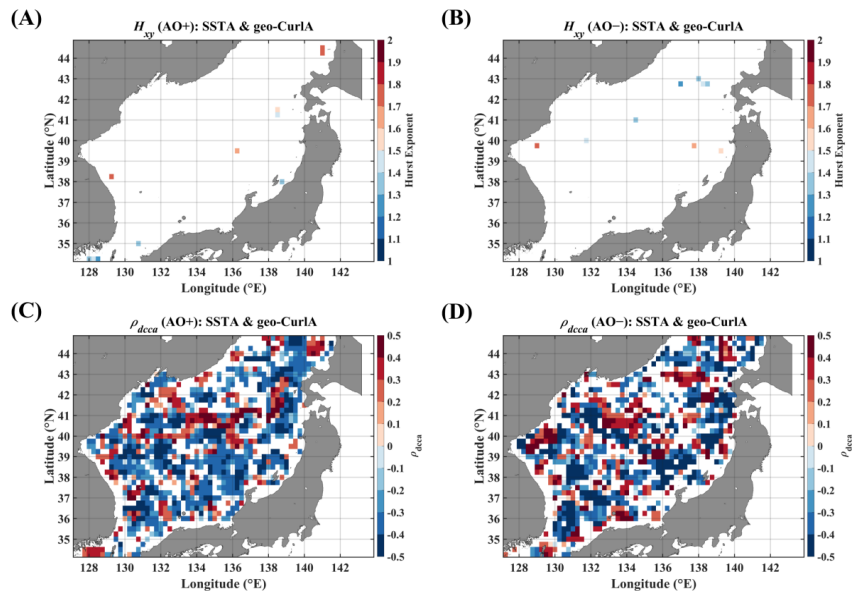

**Figure 10. SSTA↔geo-CurlA.** Same layout and significance rules as Figure 2. Curl is computed from DUACS geostrophic velocities; patchy red/blue patterns reflect geometry-dependent stirring and deformation effects.

**SSTA vs. geo-UA (Figure 11)**: $\bar{\rho}_{dcca}$ shows mixed-sign dipoles aligned with the SPF and western boundary, as expected for cross-front advection by east-west flow anomalies (the sign depends on the local meridional SST gradient). $H_{XY}$ is sporadic and confined to quasi-stationary frontal segments. AO− tends to broaden the coherent belts but does not alter the dipolar character.

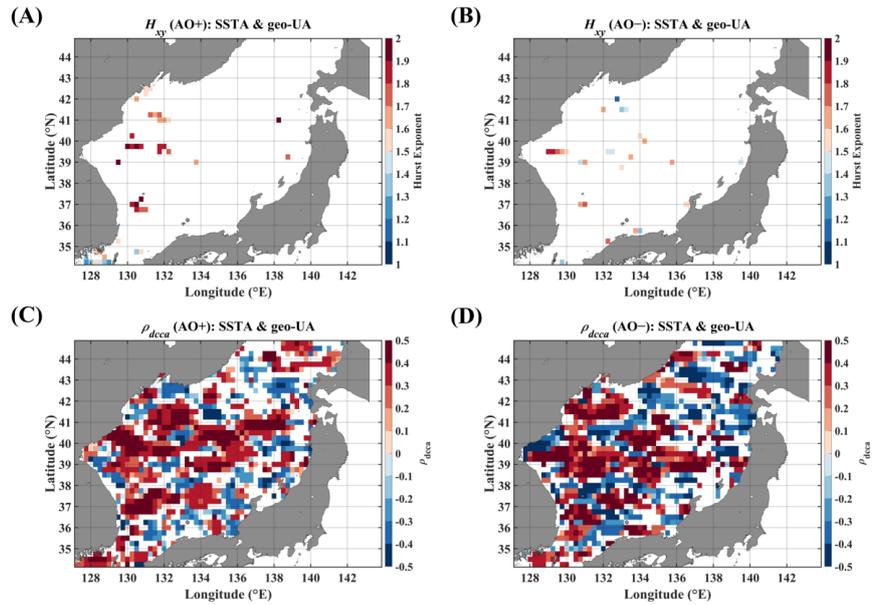

**Figure 11. SSTA↔geo-UA.** Same layout and significance rules as Figure 2. Mixed-sign $\bar{\rho}_{dcca}$ reflects cross-front advection whose sign depends on the local meridional SST gradient.

**SSTA vs. geo-VA (Figure 12)**: In contrast, geostrophic meridional anomalies present widespread positive $\bar{\rho}_{dcca}$, particularly during AO+. Significant $H_{XY}$ pixels are comparatively numerous and concentrated along the EKWC/SPF pathways, indicating that recurrent along-front advection can generate scale-invariant cross-memory at grid points embedded in persistent current routes.

Summarizing, SSHA provides the most cohesive and AO-robust positive coupling with SSTA, while geo-VA yields the clearest advective signature and the most frequent $H_{XY}$ detections along EKWC/SPF pathways. By contrast, vorticity (geo-CurlA) and zonal flow (geo-UA) yield patchy, sign-changing patterns with little cross-persistence, highlighting their sensitivity to local frontal geometry.

Taken together with Sections 3.1–3.3, the maps support a two-tier view of winter SSTA variability in the EJS: (i) mesoscale structure and along-front advection (SSHA, geo-VA) organize the persistent component and yield detectable $H_{XY}$, while (ii) synoptic winds and turbulent heat exchanges supply strong but largely non-persistent tendencies whose net impact depends on AO-phase-dependent storm tracks and their interaction with the pre-existing oceanic state.

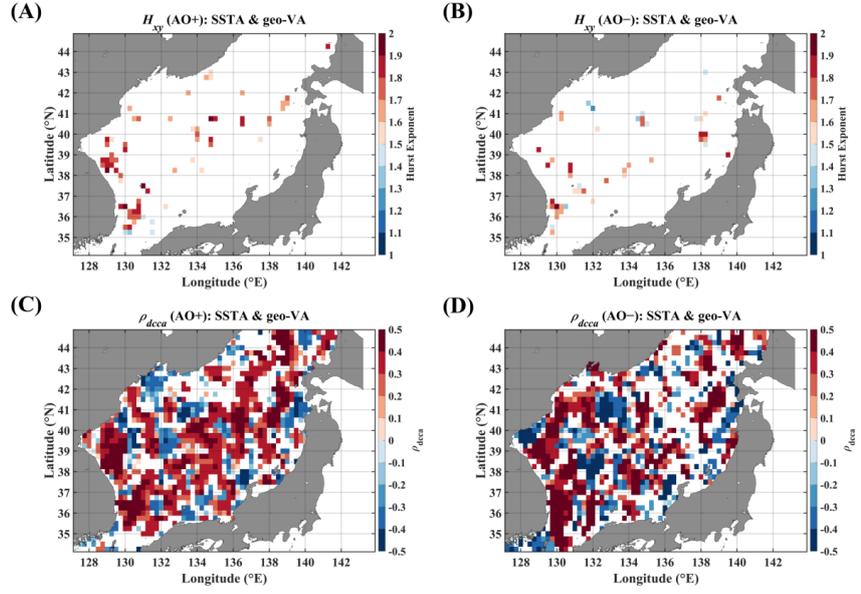

**Figure 12. SSTA↔geo-VA.** Same layout and significance rules as Figure 2. Broad positive $\bar{\rho}_{dcca}$ and relatively frequent $h_{XY}$ detections align with EKWC/SPF pathways, consistent with northward warm-advected signature.

## 4. Discussion

### 4.1. Synthesis of the main findings

A grid-point DFA/DCCA diagnostics over the EJS reveals a consistent, AO-phase-dependent hierarchy of wintertime couplings between SSTA and atmospheric, coupled-flux, and oceanic fields. First, SSTA variance and persistence ($H$) concentrate along the EKB and the SPF during AO+, with $H \approx$ 1.4–1.5 (Figure 1), indicating strong long-memory behavior in the daily anomaly field and, by implication, elevated susceptibility to persistent warm events in those corridors. Second, among atmospheric drivers, near-surface air temperature (ATMPA) exhibits basin-wide positive, scale-averaged $\bar{\rho}_{dcca}$ (5–50 days) and localized cross-persistence $H_{XY}$, whereas sea-level pressure and wind-stress curl produce patchy $\bar{\rho}_{dcca}$ and rarely yield robust $H_{XY}$ (Figures 2–4). Zonal (UA10) and meridional (VA10) winds display physically consistent signs—mostly negative and positive $\bar{\rho}_{dcca}$, respectively—with $H_{XY}$ that is sparse or confined to advective corridors. Third, SSHF and SLHF act as fast, negative feedbacks: $\bar{\rho}_{dcca} < 0$ is widespread and $H_{XY}$ is virtually absent. Finally, oceanic fields show the most cohesive coupling with SSTA: SSHA yields basin-wide positive $\bar{\rho}_{dcca}$ with $H_{XY}$ along the SPF/EKB and in the northern interior, while geostrophic V anomalies imprint the clearest advective signature and the most frequent $H_{XY}$ detections (Figures 9–12). All maps were screened with iAAFT-based Monte-Carlo p-values and BH-FDR control across space and scale.

### 4.2. Physical interpretation and AO modulation

This hierarchy aligns with known winter processes that connect AO to the East Asian circulation and air-sea exchanges. During AO+, a weakened East Asian winter monsoon favors warmer, less stormy conditions over the EJS, enhancing mixed-layer memory and along-front poleward advection along the EKWC/SPF. These conditions co-locate large SSTA variance and high $H$ (Section 3.1) and promote persistent SSTA–SSHA/geo-V co-variability (Section 3.4). During AO−, stronger outbreaks and enhanced winds reorganize wind–SST covariances, yielding sign-changing $\bar{\rho}_{dcca}$ mosaics for dynamical wind metrics (CurlTau, U10) and a broader spatial reach of positive coupling for ATMPA and V10 where advection integrates forcing into SST (Sections 3.2–3.3).

ATMPA stands out among atmospheric variables because it projects directly onto the ocean mixed layer through bulk heat exchange and stability control; hence $\bar{\rho}_{dcca}$ and episodic $H_{XY}$ in regions with strong mixed-layer memory. In contrast, SSHF/SLHF are quintessential negative feedbacks: they damp SSTA tendencies quickly, explaining the widespread $\bar{\rho}_{dcca} < 0$ and the lack of cross-scaling. Among oceanic drivers, SSHA provides a basin-coherent, AO-robust "dynamic height" control on SSTA that naturally yields positive coupling; meridional geostrophic flow (V10) maps the advective corridors where repeated along-front transport produces scale-invariant cross-memory.

### 4.3. Role of single-field memory

The appendix DFA-based $H$ maps (Figures A1–A3) clarify why some pairs develop cross-persistence while others do not. Atmospheric anomalies generally display weak persistence ($H \approx$ 0.6–1.0) with gentle spatial gradients (Figure A1), consistent with synoptic "weather" forcing that co-varies with SST without forming strong joint scaling at grid-point level. Coupled heat-flux anomalies are anti-persistent to near-white over most of the basin (Figure A2), matching their role as fast, dissipative feedbacks and explaining the absence of $H_{XY}$ with SSTA. Oceanic fields, by contrast, show high persistence ($H \approx$ 1.3–2.0) organized along the SPF, EKWC, and the northern boundary (Figure A3), mirroring the corridors of significant $H_{XY}$ and positive $\bar{\rho}_{dcca}$ in the main figures. This memory hierarchy—fluxes < atmosphere < ocean—maps directly onto the bivariate outcomes in Sections 3.2–3.4.

### 4.4. Implication for MHW susceptibility and predictability

The co-occurrence of large SSTA variance, high $H$, positive $\bar{\rho}_{dcca}$, and significant $h_{XY}$ along the EKB–SPF, especially during AO+, points to enhanced susceptibility to winter MHWs under positive AO forcing. Because the oceanic corridors that support cross-persistence are spatially stable (fronts, boundary currents), they offer subseasonal predictability windows, provided the preconditioning by mesoscale structure (SSHA, geo-VA) is monitored. By contrast, the strongly negative, memory-light coupling with turbulent fluxes limits stand-alone predictability from SSHF/SLHF anomalies.

### 4.5. Methodological considerations and limitations

DFA/DCCA are well suited to diagnose integration-like responses (for atmosphere→SST) and direct advective controls (for ocean→SST) in non-stationary daily anomalies, provided the scaling range is declared a priori and the goodness of fit is screened. Here we set a winter-appropriate 5–50-day band and enforced an $R^2$ threshold on the log–log fits; cells failing this condition were masked rather than force-tested. For statistical control, we used iAAFT surrogates (to preserve marginals and autocorrelation) and BH-FDR across space/scale, which is appropriate under the positive spatial dependence typical of geophysical fields. Observationally, DUACS geostrophic fields derive from altimetric SLA; along-track sampling and mapping can imprint anisotropy and small-scale noise. While the surrogate/FDR framework mitigates spurious detections, some fine-patch features in $H$ and $\rho_{dcca}$ may still reflect sampling and gridding choices. Finally, blanks in $H_{XY}$ maps indicate no reliable power-law joint scaling; in such cases, scale-specific $\rho_{dcca}$ becomes the appropriate metric of local co-variability.

**Author Contributions:** Conceptualization, G.L. and J.-J.P.; methodology, G.L.; validation, G.L.; formal analysis, G.L.; investigation, G.L.; data curation, G.L.; writing—original draft preparation, G.L.; writing—review and editing, G.L. and J.-J.P.; visualization, G.L.; supervision, J.-J.P.; project administration, G.L. and J.-J.P.; funding acquisition, G.L. and J.-J.P. All authors have read and agreed to the published version of the manuscript.

**Funding:** This research was supported by the Korea Institute of Marine Science & Technology Promotion (KIMST) funded by the Ministry of Oceans and Fisheries (RS-2023-00256005). This work was also supported by the National Research Foundation of Korea (NRF) grant funded by the Korean government (MSIT) (RS-2022-NR069134). G.L. was supported by the National Research Foundation of Korea (NRF) grant funded by the Korean government (MSIT) (RS-2024-00507484).

**Data Availability Statement:** The monthly Arctic Oscillation (AO) index was obtained from the NOAA Climate Prediction Center (CPC) (available at the CPC website at https://www.cpc.ncep.noaa.gov/; accessed on 10 April 2024).

The daily NOAA Optimum Interpolation SST v2.1 (OISST) dataset is publicly available from the NOAA Physical Sciences Laboratory (PSL) at https://psl.noaa.gov/data/gridded/data.noaa.oisst.v2.highres.html (accessed on 4 April 2023).

The ERA5 reanalysis single-level variables (10m zonal/meridional winds, sea level pressure, 2m atmospheric temperature, surface sensible and latent heat fluxes) are publicly available from the Copernicus Climate Data Store (CDS) at https://cds.climate.copernicus.eu/cdsapp#!/dataset/reanalysis-era5-single-levels (accessed on 10 April 2025).

Sea-level anomaly and derived geostrophic products (DUACS, 1/4°) are available from the Copernicus Climate Change Service (C3S) Climate Data Store (DOI: 10.24381/cds.4c328c78 (accessed on 4 April 2023).

**Acknowledgments:** NOAA OISST v2.1 data were provided by the NOAA PSL, Boulder, CO, USA (https://psl.noaa.gov (accessed on 4 April 2023). We also acknowledge ECMWF for ERA5 reanalysis and the Copernicus Climate Change Service (C3S) for DUACS altimeter products.

**Conflicts of Interest:** The authors declare no conflict of interest.

## Abbreviations

The following abbreviations are used in this manuscript:

**AO**: Arctic Oscillation (AO+ / AO− denote positive/negative phases)

**ATMP, ATMPA**: 2-m air temperature; its anomaly

**BH**: Benjamini–Hochberg (false discovery rate procedure)

**C3S**: Copernicus Climate Change Service

**CDS**: Copernicus Climate Data Store

**CPC**: Climate Prediction Center (NOAA)

**DCCA**: Detrended Cross-Correlation Analysis

**DFA**: Detrended Fluctuation Analysis

**DUACS**: Data Unification and Altimeter Combination System

**EAWM**: East Asian Winter Monsoon

**EJS**: East/Japan Sea

**EKWC**: East Korea Warm Current

**EKB**: East Korean Bay

**ERA5**: ECMWF Reanalysis v5

**FDR**: False Discovery Rate

**iAAFT**: iterative Amplitude-Adjusted Fourier Transform (surrogates)

**JFM**: January–February–March (winter season)

**MHW(s)**: Marine Heatwave(s)

**SLP, SLPA**: Sea-level pressure; its anomaly

**SLHF, SLHFA**: Surface latent heat flux; its anomaly

**SSH, SSHA**: Sea surface height; its anomaly

**SSHF, SSHFA**: Surface sensible heat flux; its anomaly

**SST, SSTA**: Sea surface temperature; its anomaly

**SPF**: Subpolar Front

**TWC**: Tsushima Warm Current

**U10/V10; UA10/VA10**: 10-m zonal/meridional winds and their anomalies

**CurlTau, CurlTauA**: Wind-stress curl and its anomaly

**geo-U/geo-V; geo-UA/geo-VA**: Geostrophic zonal/meridional currents and their anomalies

**geo-Curl; geo-CurlA**: Geostrophic current curl and its anomaly

**Appendix A**

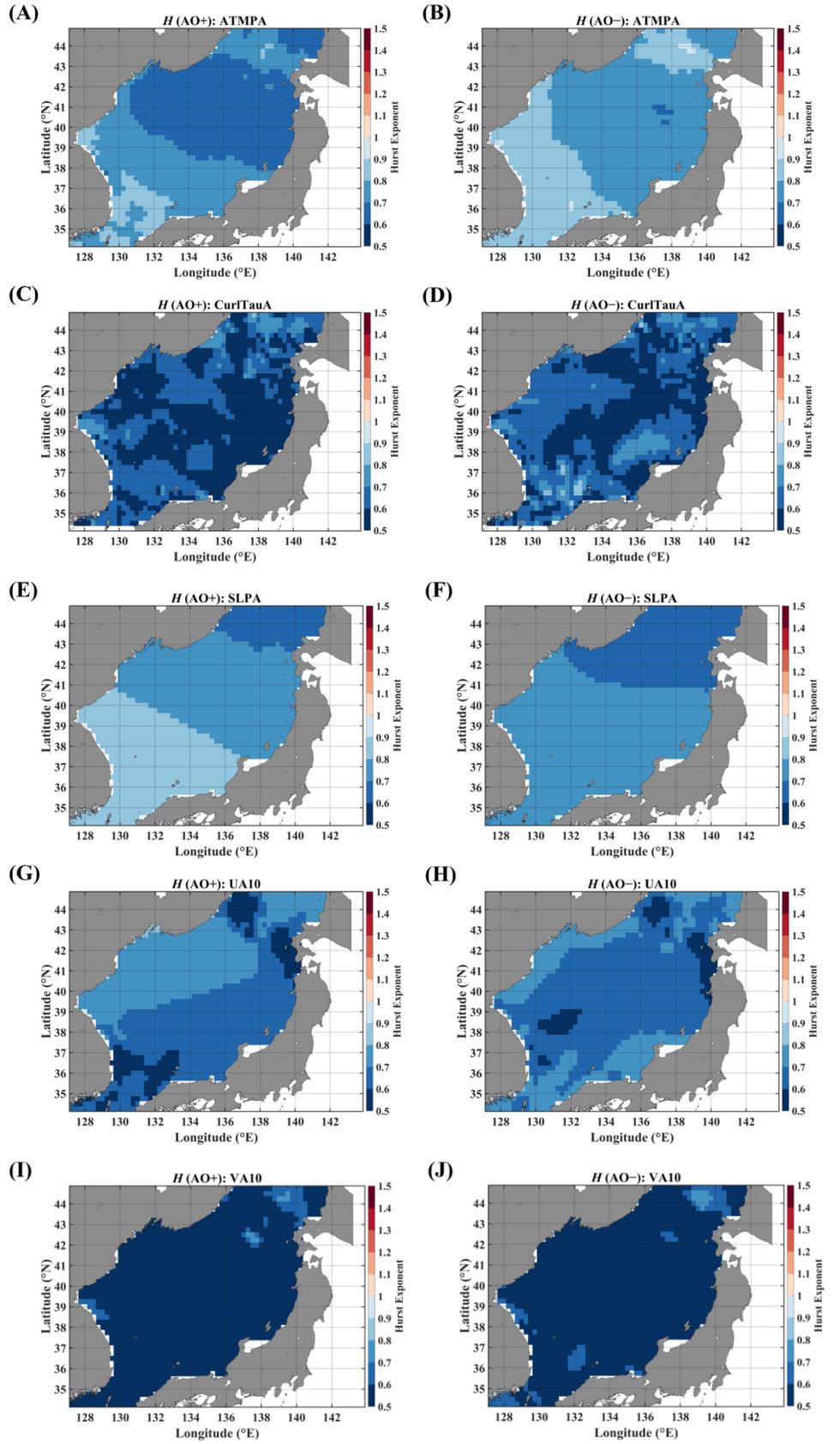

**Figure A1**. DFA Hurst exponent ($H$) of atmospheric anomalies under AO+ (left) and AO− (right): (A,B) ATMPA; (C,D) CurlTauA; (E,F) SLPA; (G,H) UA10; (I,J) VA10. $H$ is estimated from linear fits of $\log F(s)$ vs. $\log s$ over $s$= 5–50 days.

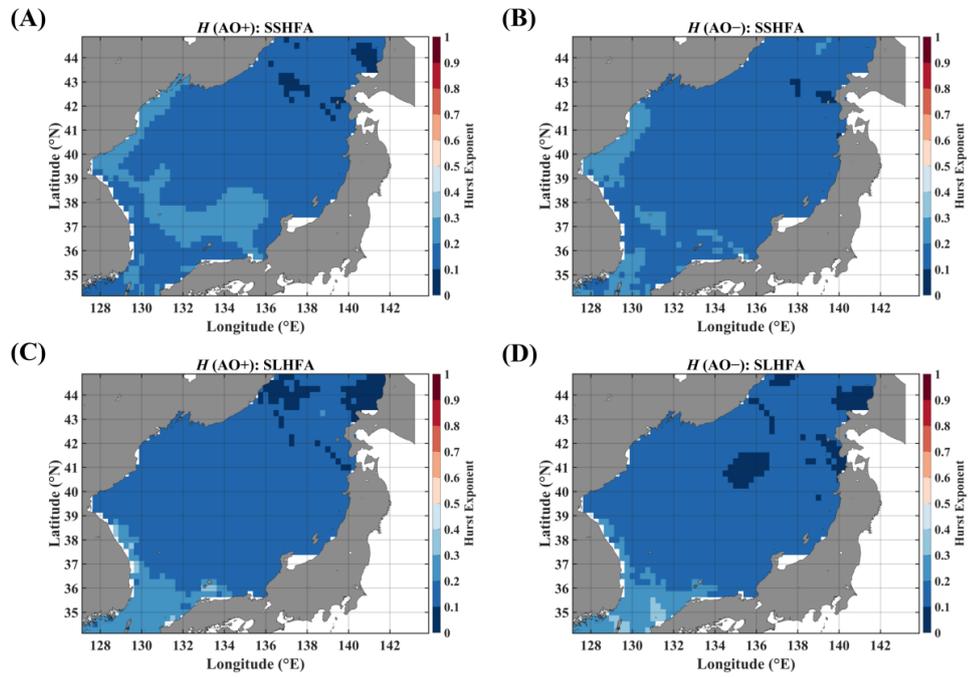

**Figure A2**. DFA $H$ of coupled heat-flux anomalies: (A,B) SSHFA; (C,D) SLHFA, shown for AO+ (left) and AO− (right). The 0–1 color scale emphasizes anti-persistent to near-white behavior ($H$ < 0.5).

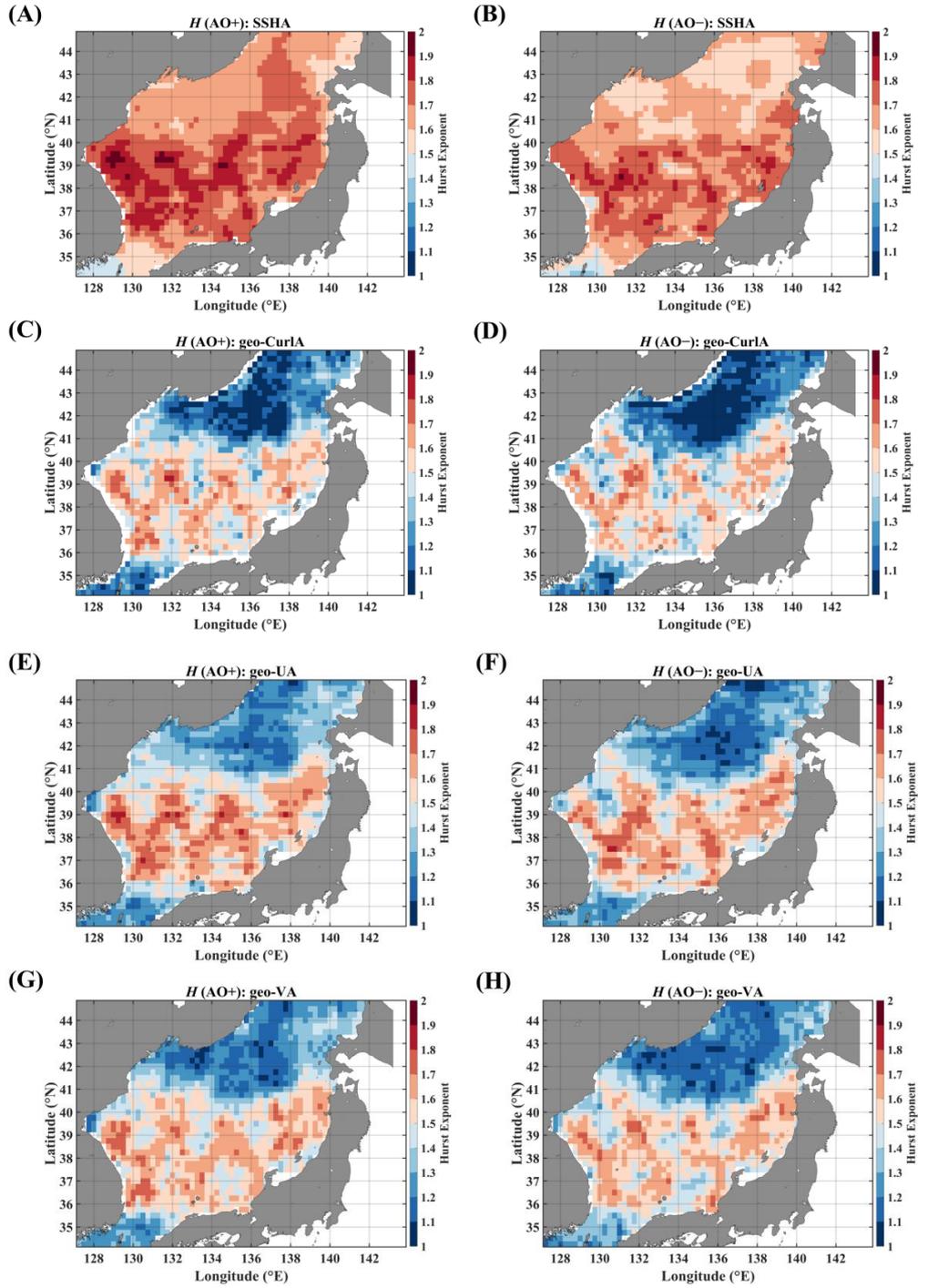

**Figure A3**. DFA $H$ of oceanic anomalies: (A,B) SSHA; (C,D) geostrophic current curl anomaly; (E,F) geostrophic UA; (G,H) geostrophic VA, under AO+ (left) and AO− (right). High persistence ($H \approx 1.3–2.0$) is organized along the SPF/EKWC and northern boundary, consistent with corridors of significant $H_{XY}$ and positive $\bar{\rho}_{dcca}$ in the main figures.